\newcommand{\beq}{  \begin{eqnarray}}
\newcommand{\eeq}{  \end{eqnarray}}

\documentclass[pra,twocolumn,shopacs,preprintnumbers,asmath,amssymb]{revtex4-1}

\usepackage{amsmath,amssymb}
\usepackage{subfig}
\usepackage{graphicx}% Include figure files
\usepackage{float}
\usepackage{dcolumn}% Align table columns on decimal point
\usepackage{bm}% bold math

\begin{document}

\title{The Molecular Aharonov-Bohm Effect Redux}
\email{bernard@physics.unlv.edu}

\author{B. Zygelman}

\affiliation{Department of Physics and Astronomy, University of Nevada, Las Vegas, USA}

\begin{abstract}
 A solvable molecular collision model that
predicts  Aharonov-Bohm (AB) like scattering in the adiabatic approximation is introduced. 
For it, we propagate coupled channel wave packets without resorting to
a Born-Oppenheimer (BO) approximation. In those, exact, solutions we find evidence of topological phase dislocation lines that
are independent of the collision energy and provide definitive signatures of AB-like scattering. The results of these simulations
contrast with the conclusions of a recent study that suggests survival of the
molecular Aharonov-Bohm (MAB) effect only in the adiabatic limit in which the nuclear reduced mass $\mu \rightarrow \infty$.
 We discuss generalizations of this model and consider
possible screening of the Mead-Truhlar vector potential by the presence of multiple conical intersections (CI). 
We demonstrate that the Wilson loop phase
integral has the value $-1$ if it encloses an odd-number of CI's, and takes the value $+1$ for an even number.
Within the scope of this model, we investigate the ultra-cold limit of scattering solutions in the presence of a conical intersection and comment
on the relevance of Wigner threshold behavior for s-wave scattering.
\end{abstract}

\maketitle
\section{Introduction}

The Aharonov-Bohm effect\cite{ab59,AB49} describes the behavior of a charged particle in the presence of
a gauge vector potential that does not impress a Lorentz force on the particle, but nevertheless exerts a profound influence on its scattering properties.
Importantly, it is a topological effect and, as such, has served as a template in understanding the behavior of exotic forms of quantum matter including anyons\cite{Wilczek82}, quantum Hall systems\cite{FQH83}, and topological insulators\cite{TOPI2010}.

In molecular physics, an analog of the AB effect was discovered by Mead and Truhlar\cite{mead76} in their analysis of polyatomic systems whose electronic Born-Oppenheimer surfaces posses a conical intersection.  They showed how, in the ground BO surface that shares a CI with an excited 
electronic BO state, the motion of atoms are minimally coupled to an effective vector potential similar to that
which describes a magnetic flux tube
\beq
{\bm A}  =   {\bm {\hat \phi} }   \frac{\Phi} {2 \pi R}.  \label{1.6}
\eeq
Here $ \Phi$ is the magnetic flux enclosed by the (infinitesimal) tube running along the z-axis in a cylindrical coordinate system in which
$  {\bm {\hat \phi} } $ is the azimuthal unit vector and $R$ is the distance from the flux tube. 
The scattering amplitude is proportional to the enclosed magnetic flux  provided that $  \Phi/2\pi  \neq n$, where $n$ is an integer.
In the Mead-Truhlar analysis the molecular reaction coordinates are coupled to this vector potential for the case $\Phi =\pi$ which
we henceforth label as $ {\bm A}_{MAB} $. It arises 
due to the properties of the BO electronic wavefunctions near a CI and which is the locus of an effective flux tube.  
According to AB theory, ${\bm A}_{MAB}$  should give rise to topological effects in a reactive scattering setting and has therefore been called the molecular Aharonov -Bohm effect (MAB).  

The bound state Aharonov-Bohm effect\cite{peshkin,mead80} describes the shift in the eigenenergies
of a bound system in the vicinity of such a flux tube. It is in this context
 that the MAB effect was first observed in a laboratory setting \cite{Zwanzig,Demtroder}.
In Jahn-Teller (JT)\cite{Zwanzig,zwanzig87} systems, in which the atoms are subjected to a bounding 
``Mexican hat'' scalar potential circumscribing the conical intersection vibrational eigenvalues are shifted, 
and have been detected in spectroscopic studies

Despite the success of the gauge paradigm to predict bound state energy shifts in Jahn-Teller systems, 
more than a quarter  century of effort in both theoretical and experimental arenas have failed to provide a clear signature
for the molecular AB effect in a reactive scattering scenario.  Early theoretical studies\cite{Wu93}  of the ro-vibrational product
distribution in reactive scattering of the ${\rm H} + {\rm H}_{2} $ system, included the MAB effect and promised to resolve\cite{Levii93} existing discrepancies between experiment\cite{zare92} and theory.  Subsequently, additional theoretical efforts and experiments showed that the issue is not so clear cut\cite{wrede97,ken00,marcos1,marcos05,zare2014}. For example,  more recent experimental measurements\cite{wrede97} for product state distributions
are in excellent agreement with calculations that omit the MAB effect\cite{marcos1}.

Theoretical and numerical studies of the MAB effect in reactive scattering systems 
fall into two categories. In the first, an adiabatic approximation
is employed and non-adiabatic couplings to states other than the ground electronic states are ignored.
 The ground state amplitude is minimally coupled to ${\bm A}_{MAB}$, or boundary conditions on the vibronic amplitude are imposed so that the product of vibronic
and the ground (multi-valued) electronic amplitudes is single-valued.  In the second category, a two-state approximation which includes both the ground and excited electronic 
states that share a CI are incorporated. That approach leads to a pair of coupled Schroedinger-like equations,
and it has been argued that, in this case, it is not necessary to include ${\bm A}_{MAB}$. 
In a recent numerical study\cite{gross15}  adiabaticity  was relaxed in a so-called exact treatment of a model system that possesses a CI. 
In those calculations it was found
that the MAB effect survives only in the limit in which the effective mass for the vibronic motion tends to infinity. 
That result suggests that evidence of a MAB effect, cited in studies based on the BO approximation, may be an artifact of the latter. 
In calculations involving realistic systems, issues such as the accuracy of potential
surfaces, inclusion of rotational and other couplings, accommodation of realistic asymptotic boundary conditions, arise
and  need to be addressed in a rigorous and un-ambiguous manner.  For example, it is known\cite{zyg87a,moo86} that rotational couplings
can also be described by an effective vector gauge potential. 
The latter may  lead to phase holonomies that interfere with those generated by ${\bm A}_{MAB}$. 

Because of the cited discrepancy and the lack of a clear experimental signature, 
we are motivated to re-consider the question; is the MAB effect an artifact of an adiabatic approximation?
Definitive predictions and validation requires a system
that exhibits the requisite complexity, i.e. it possesses a CI, but at the same time it must be simple enough
to allow accurate numerical solution. In the discussion below, we introduce such a model and offer exact numerical solutions that are
not limited by adiabatic assumptions. Though it does not describe a realistic molecular system, it shares essential features of the latter and 
predictions gleaned from it enhance our understanding of MAB phenomena in general.
  Before introducing and solving the proposed model, we first review classical pure
AB scattering\cite{ab59}, and sharpen our understanding of what it means for the latter to be a topological effect.

\subsection{Classical AB scattering}
Consider an incoming packet with de-Broglie wavelength $\lambda$ impinging on the magnetic flux tube described by  vector potential 
Eq.(\ref{1.6} ). The particle is scattered by it  and , in a time independent description,  the differential scattering cross section (per unit length) is\cite{ab59},
\beq
\lambda \, \frac{d\sigma}{d \theta} =  \frac{\sin^{2} \pi \alpha}{\cos^{2} \theta/2 } 
\label{1.6a}
\eeq
where $\alpha \equiv e \Phi/h$, and  $\theta$ is the angle measured from the incident  ($+x$ direction)  flux.
The right hand side of Eq. (\ref{1.6a}) is independent of the collision energy and is a reflection of the topological nature of AB scattering.
Consider now the case where the flux tube is embedded along the axis of an impenetrable cylinder of radius $a$. We then have a scenario in which
the scattering amplitude of the short range potential (of the cylinder)  interferes with that of the long-range nature of Eq. (\ref{1.6}). The total
wave functions is now described by\cite{Berry80}
\beq
&& \psi(r,\theta) = \psi_{AB}(r,\theta) - \psi_{a} (r,\theta)  \nonumber \\
&& \psi_{a} (r,\theta)=  \nonumber \\
&& \sum_{m } (-i)^{|m-\alpha |}  \exp(i m \theta) \frac{J_{| m-\alpha |} (k a)}{  H^{(1)}_{| m-\alpha |} (k a)} \, 
 H^{(1)}_{| m-\alpha |} (k r) 
\label{1.6c}
\eeq
\begin{figure}[H]
\centering
\includegraphics[width=0.8 \linewidth]{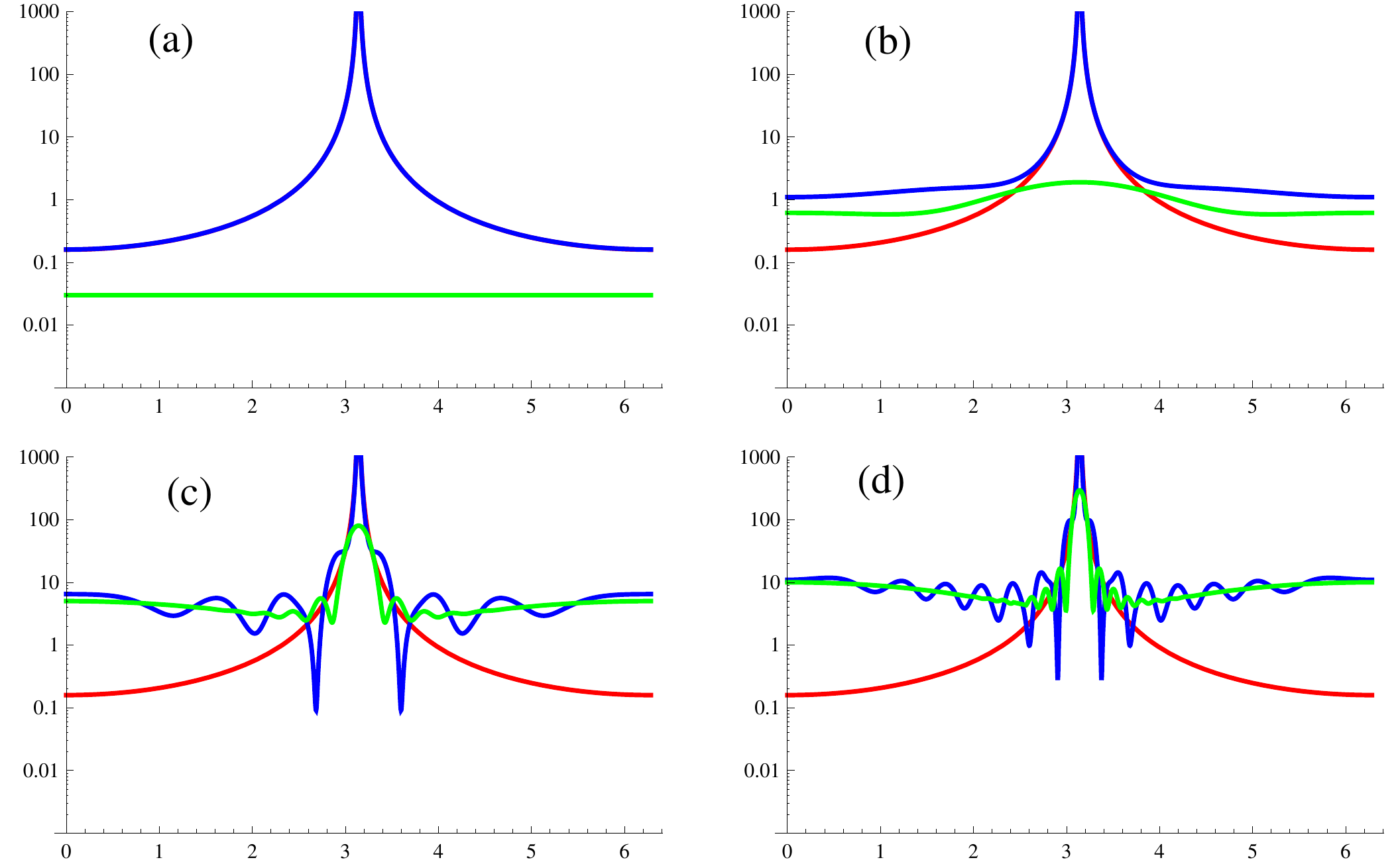}
\caption{\label{fig:fig1}  Plot of $ k  \, d \sigma/d\theta $, where the abscissa is the scattering angle $0 \le \theta < 2 \pi $. 
Green lines correspond to scattering by the cylinder only, red lines identify pure AB scattering, blue lines denotes scattering by the cylinder including
${\bm A}_{MAB}$.  
The panels (a)-(d) correspond to wave numbers $k=0.01,0.1,1,10 $ respectively.}
\end{figure}
\noindent
where $\psi_{AB}$ is the AB wavefunction and the sum extends over all negative and positive integers $m$.
For the case $\alpha=1/2$,  which corresponds to the MAB value for $\Phi$,  $\psi_{AB}$ has the analytic form\cite{ab59,Berry80}
\beq
&& \psi_{AB}  =  -\exp(i \phi/2) \exp(-i k \rho \cos \phi )
 \times \nonumber \\ 
&&  Erf(\exp( i 3\pi/4) \sqrt{2 k \rho } \cos(\phi/2) )
\label{1.6d}
\eeq
where $ \tan\phi= x/y$,  $k^{2}/2$ is the collision energy and $ 0 \leq |\phi| \leq \pi $. 
At large $ r = \sqrt{x^{2}+y^{2}} $ it describes an incident wave, approaching from the positive $x$  direction, 
\beq
\psi_{inc} = \exp(-i k x) \exp(i \phi/2). 
\label{1.6e}
\eeq
Though the incident wave appears to be
multi-valued, the total amplitude is single-valued\cite{Berry80}.

In Figure (\ref{fig:fig1}) we plot the differential cross sections, as a function of the scattering angle $\theta$, for various values of incoming wavenumbers $k$ for
the cases (i) scattering by an impenetrable cylinder without the presence of a flux tube, (ii) pure AB scattering (no short range potential), (iii) the case described
by the wave function Eq. (\ref{1.6c}). The green lines in panels (a)-(d) give $k \,d\sigma/d\theta$ for case (i).
At very low collision energies, corresponding to the s-wave scattering limit, $ d\sigma/d\theta$ tends to a constant for all values of $\theta$. At higher
energies the cross sections exhibit pronounced structures and approaches, at larger $\theta$, the classical value (shown by the dotted black line in panel (d)).
The strong forward scattering peak is indicative of wave interference (i.e. the 1D analog of the Poisson spot\cite{zyg15} ).  The red lines  are plots  of Eq. ({\ref{1.6a}) and
are independent of the wave number $k$. In panel (a), corresponding to $k=0.01$, the red line is not visible as the blue line overlaps it and represents scattering by an AB flux tube
enclosed by an impenetrable cylinder of radius $a=1$. The panels illustrate the interference between the short range interaction
with the cylinder and the long range coupling with gauge field ${\bm A}_{MAB}$ for various collision energies. As $k$ increases we find that interference effects
begin to wash out (except for a small region toward the forward scattering angle)  contributions that arise from ${\bm A}_{MAB}$.
In panel (d)  the cross sections, given by the blue line, approach that of scattering
by the short range interaction only (i.e. the green line).  

Though the influence of ${\bm A}_{MAB}$  in Fig. (\ref{fig:fig1}) is evident, especially at lower collisions energies, interference 
with the scattered waves of the short range potential does not offer a compelling demonstration of the inherent topological nature of the
former as the collision energy is varied. Inspection of $d\sigma/d\theta$ (for $\theta \neq 0 $) is not an ideal indicator of hidden
topological order. Instead, we follow the strategy of  Ref. \cite{Berry80} and study features that are topological invariants. 
In Fig (\ref{fig:fig2}) we plot the imaginary part of the total wave function $\psi(R,\theta) $, given in Eq. (\ref{1.6c}). It is the wave function
that describes scattering by a short range potential that includes the AB vector potential. In that figure the abscissa represents the $x$ axis, with the incident
wave being scattered by the cylinder, shown as the white circle, at the origin. The horizontal axis defines the $y$ axis. The leftmost panel
corresponds to the wavenumber $k=0.01$, the middle panel to $k=1$, and the rightmost panel to $k=10$.  
In those figures we find, as expected, complex interference structures between incident and scattered waves as the collision energy is varied. However, we also
note prominent a phase dislocation line (which corresponds to a nodal line of the total amplitude) that originate at the origin and extends along the negative $x$ axis. These features, unlike $d\sigma/d\omega$,
remain fixed as the collision energy is varied. It was pointed out in Ref. \cite{Berry80} that such phase dislocation lines are
  topological invariants. That dislocation line extends along the negative $x$ axis and is clearly visible in Fig. (\ref{fig:fig2}).
\begin{figure}[H]
\centering
\includegraphics[width=0.9 \linewidth]{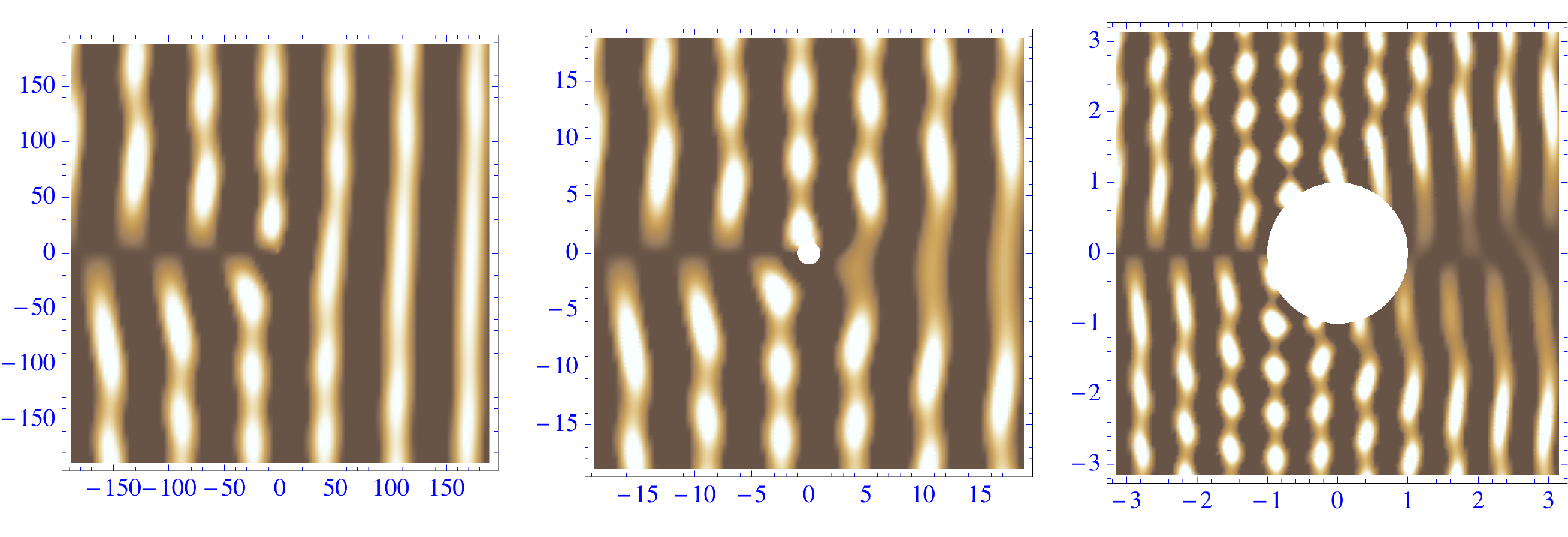}
\caption{\label{fig:fig2} Density plots of the imaginary part of $\psi(R,\theta)$ as function of $x$ (abscissa) and $y$ (ordinate). The panels from left to right correspond
to wave numbers $k=0.01, 1, 10 $ respectively. The incident wave approaches from the right. The white disk at the center represents an impenetrable cylinder of radius $a=1$. }
\end{figure}
In the calculations presented here we consider a molecular collision model that is expected to demonstrate Mead-Truhlar phase holonomy
in the adiabatic (i.e. the BO) approximation. However we relax adiabaticity and, within the scope of this model, proceed to produce exact numerical solutions for the scattering problem. We use the resulting solutions to investigate wether the aforementioned, topological,  AB-like phase discontinuities arise and persist as the collision energy is varied. 

\subsection{Gauge potentials induced by conical intersections}
Consider a tri-atomic system that possesses a conical intersection at the origin of a parameter space  spanned by a set
of nuclear internal coordinates  $x,y$. Typically,  they represent various linear combinations of the squares of inter-nuclear distances between the three nuclei\cite{mead76} in a planar configuration. In this coordinate system the azimuthal angle $\phi$ is called the pseudo-rotation and $\rho=\sqrt{x^{2}+y^{2}} $ measures distortions from an equilateral triangle
configuration of nuclei. We describe the system by an amplitude $\psi(x,y,{\bm r})$ where ${\bm r}$ are electronic coordinates.  
Once the electronic, or fast, coordinates are integrated out  the adiabatic Hamiltonian is a truncated two-dimensional Hilbert space operator,  which in the vicinity of the intersection is given by\cite{teller}
\beq
H_{ad}=
\left(
\begin{array}{cc}
  x  &  y    \\
  y  &  -x   \\
\end{array}
\right).
\label{2.0}
\eeq
The eigenvalues of $H_{ad}$ are $\pm \sqrt{x^{2}+y^{2}}$ and correspond to first excited and ground states, respectively,
of the electronic Hamiltonian. The eigenstates of $H_{ad}(x,y)$ are parameterized by the nuclear coordinates and form 
the adiabatic basis for the PSS expansion, which for this model is complete. 
Figure (\ref{fig:fig1}) illustrates a typical conical intersection located at the origin of our
coordinate system.  

Because $H_{ad}$ is real, Longuet-Higgins and Herzberg\cite{Herzberg} constrained its eigenstates
 to be real-valued and found
\beq
 && |\Phi_{g}  \rangle = \left ( \begin{array}{c}    
                            - \sin\phi/2 \\
                             \cos\phi/2 
                             \end{array}  \right )     =   {\tilde U}(\phi) |g \rangle  \nonumber \\
 && {\tilde U}(\phi) =  \left ( \begin{array}{cc}    
                            \cos\phi/2  & -\sin\phi/2 \\
                            \sin\phi/2  & \cos\phi/2 
                             \end{array}  \right )   \quad  \nonumber \\                        
 && |g \rangle = \left ( \begin{array}{c}    
                            0 \\
                             1
                             \end{array}  \right ),     \quad  |e \rangle = \left ( \begin{array}{c}    
                            1 \\
                             0
                             \end{array}  \right ), 
 \label{2.1}                             
\eeq
where $|\Phi_{g} \rangle $ is the ground adiabatic electronic state.
They noted that it is multi-valued, 
as its value changes sign in traversing a circuit from $\phi=0$ to $\phi=2\pi$. The total system amplitude $\psi$
must be single valued and so in a Born-Oppenheimer approximation in which $  \psi= F(x,y) |\Phi_{g} \rangle $, the vibronic amplitude $F(x,y)$ must undergo a compensating sign change. That argument inspired Mead and Truhlar\cite{mead76} to invoke the minimal
coupling of the vibronic motion with the vector potential $ {\bm A}_{MAB}$.
\begin{figure}[ht]
\centering
\includegraphics[width=0.6 \linewidth]{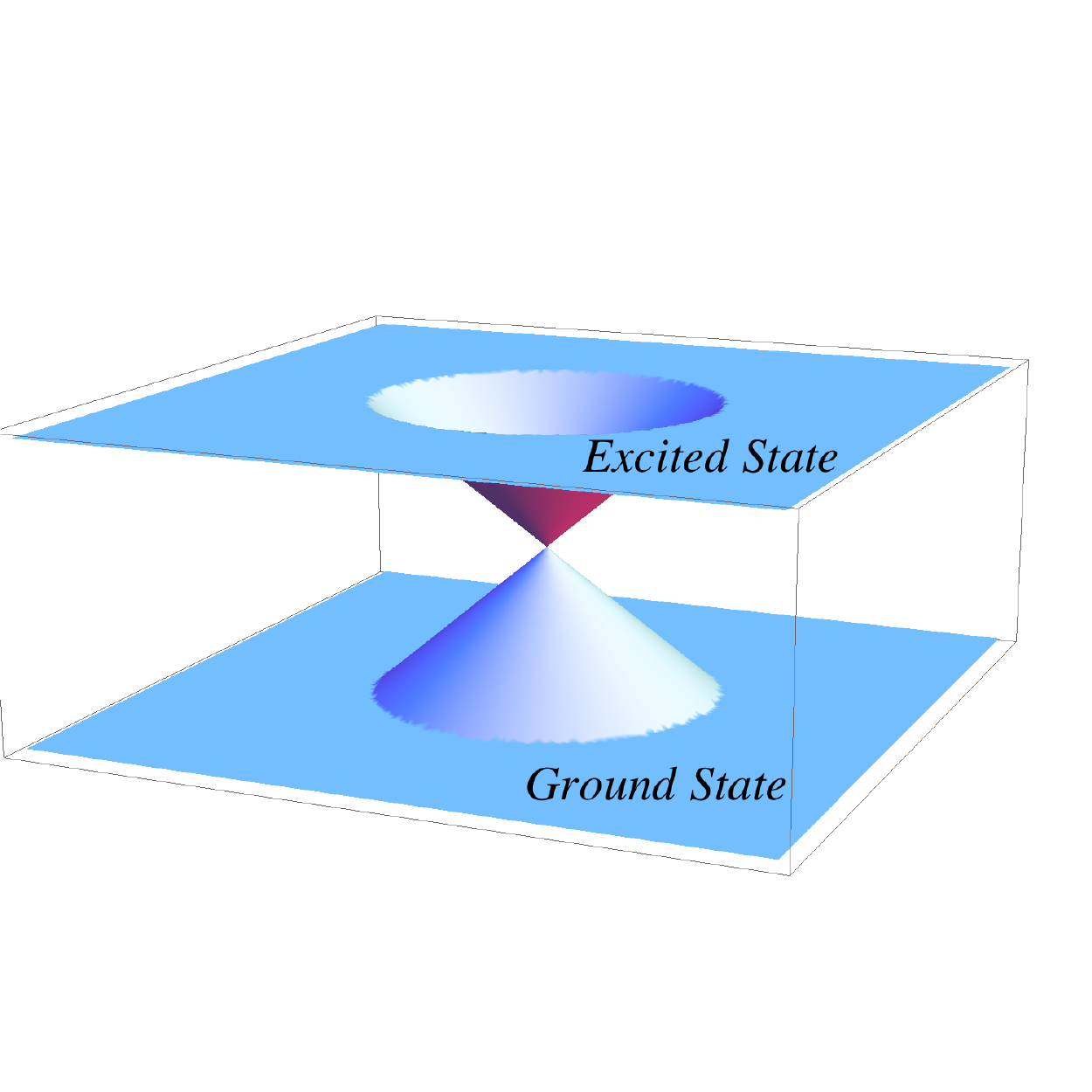}
\caption{\label{fig:fig3} (Color online). Illustration of a conical intersection between the ground BO surface and an excited 
electronic BO surface.}
\end{figure}

In this analysis we offer an alternative tack to that summarized above.
Proceeding along the lines outlined in Ref. \cite{zyg87a}, which provides a general gauge theory setting
for the PSS equations,  we do not 
constrain the phases of the BO expansion basis $\Phi_{n}({\bm R},{\bm r})$ but we do require them
to be single valued for all ${\bm R}$.  Applying this prescription to the model adiabatic Hamiltonian
Eq. (\ref{2.0}) we find that  
\beq
&& H_{ad} = U_{c}(\phi) H_{BO} U_{c}^{\dag}(\phi)  \nonumber \\
\nonumber \\
&&H_{BO} = \left ( \begin{array}{cc}    
                            \sqrt{x^{2}+y^{2}} & 0  \\
                             0 &  -\sqrt{x^{2}+y^{2}}
                             \end{array}  \right )  
\label{2.3}
\eeq
where 
\beq
&& U_{c}(\phi) = \exp(-i \sigma_{2} \phi/2) \exp(i \sigma_{3}\phi/2) = \nonumber \\
\nonumber \\
&&    \left(
\begin{array}{cc}
 e^{\frac{i \phi }{2}} \cos \left(\frac{\phi }{2}\right) & -e^{-\frac{i \phi
   }{2}} \sin \left(\frac{\phi }{2}\right) \\
 e^{\frac{i \phi }{2}} \sin \left(\frac{\phi }{2}\right) & e^{-\frac{i \phi
   }{2}} \cos \left(\frac{\phi }{2}\right) \\
\end{array}
\right).  
\label{2.4}                        
\eeq
Unlike the operator $ {\tilde U}(\phi)$ given in Eq. (\ref{2.1}),  which undergoes a sign change as $\phi$ ranges
from $0$ to $2\pi$, $U_{c}(\phi)$  is single valued for all $\phi$, excluding the origin, i.e. $U(\phi+2\pi)=U(\phi)$. 
So we employ the {\bf single-valued } eigenstates $ U_{c}(\phi)| g \rangle $, and $ U_{c}(\phi)| e \rangle $ 
in our the PSS expansion.  Taking the vibronic kinetic energy operator to have the form
$$ H_{KE} =-\frac{\hbar^{2}}{2 \mu} \Bigl( \frac{\partial^{2}}{\partial x^{2}} + \frac{\partial^{2}}{\partial y^{2}} \Bigr ) $$ where 
$\mu$ is a reduced atomic mass.  We obtain (e.g see Ref.\cite{zyg15})
\beq
 -\, \frac{\hbar^{2}}{2 \mu} \Bigl ( {\bm \nabla} - i {\bm A} \Bigr )^{2} F({\bm R}) + 
{ V}({\bm R}) F({\bm R}) = E F({\bm R}). 
\label{2.5}
\eeq
$F({\bf R})$ is a column vector whose two entries are the ground and excited state adiabatic vibronic amplitudes,
 $V({\bm R} )=H_{BO}(x,y) $  and 
${\bm A}$ is the matrix gauge  potential
\beq
 {\bm A} = i U_{c}^{\dag} {\bm {\nabla}} U_{c} = 
\frac{ \bm {\hat \phi }}{2\rho}  \left(
\begin{array}{cc}
   -1 &  -i \exp(-i \phi)   \\
    i \exp(i \phi)   &  1  
\end{array}
\right).
\label{2.6}
\eeq

Because the PSS expansion basis is complete,  Eq. (\ref{2.5}) is equivalent to the coupled equations
obtained in the diabatic basis set i.e., the amplitude $G({\bm R}) = U_{c}(\phi) F({\bm R})$ satisfies
\beq
 -\, \frac{\hbar^{2}}{2 \mu} {\bm \nabla}^{2} G({\bm R}) + H_{ad}(x,y) G({\bm R}) = E G({\bm R})
\label{2.7}
\eeq
where $H_{ad}(x,y)$ is given by Eq. ({\ref{2.0}). 

In order to study collision phenomena, which require  in-coming and out-going asymptotic packets,
we introduce a slightly modified version of the LHH model, and in which
the diabatic coupling is given by
\beq 
&& H'_{ad}(x,y) = \Xi (\rho,\rho_{0})  \, \left ( \begin{array}{cc}
   x &  y  \\
    y   &  -x  
\end{array} \right ) \nonumber \\
&&   \frac{\Xi}{\Delta} \equiv \frac{\theta(\rho-\rho_{0})}{\rho} +  \frac{ \theta(\rho_{0} -\rho)}{\rho_{0}} 
\label{2.8}
\eeq
where $ \rho_{0}, \Delta $ are constants and $\theta$ is the Heaviside function.  The eigenvalues of $ H'_{ad}(x,y)$ are
\beq
&& \pm \, \Delta \quad {\rm for}  \, \rho > \rho_{0}  \nonumber \\
&& \pm \, \frac{\Delta}{\rho_{0}}  \, \sqrt{x^{2}+y^{2}}  \quad {\rm for} \, \rho <= \rho_{0}.
\label{2.9}
\eeq
The BO eigenvalues of $H'_{ad}(x,y)$ describe a conical intersection, centered on the origin but in the region $\sqrt{x^2+y^2} > \rho_{0}$ 
it consists of two flat surfaces separated by an energy gap $2 \Delta$. The latter feature allows asymptotic incoming and outgoing
scattering states. The BO surfaces, defined  by Eq. (\ref{2.9}), are illustrated in Figure (\ref{fig:fig3}). 
We now study the behavior of an asymptotic ``free'' wave packet as it approaches in the ground BO state and is scattered by the CI
centered at the origin. To that end we employ the split operator method\cite{fleck} to propagate the wave packet. 
At some initial time $t_{0}$ we place a packet in the ground BO state  shown in panel (a) of 
Figure (\ref{fig:fig4}). The mean velocity of the packet is chosen so that it approaches the origin at subsequent times $t>t_{0}$. 
Panels (b) and (c) of Figure
 (\ref{fig:fig4}) illustrate it's time development as it approaches the CI, diffracts about it, and eventually continues, shown in panel (d), 
as a free, scattered, packet at the post collision time $t_{f}$.  The detailed discussion of this calculation is summarized below.
\begin{figure}[H]
\centering
\includegraphics[width=0.6 \linewidth]{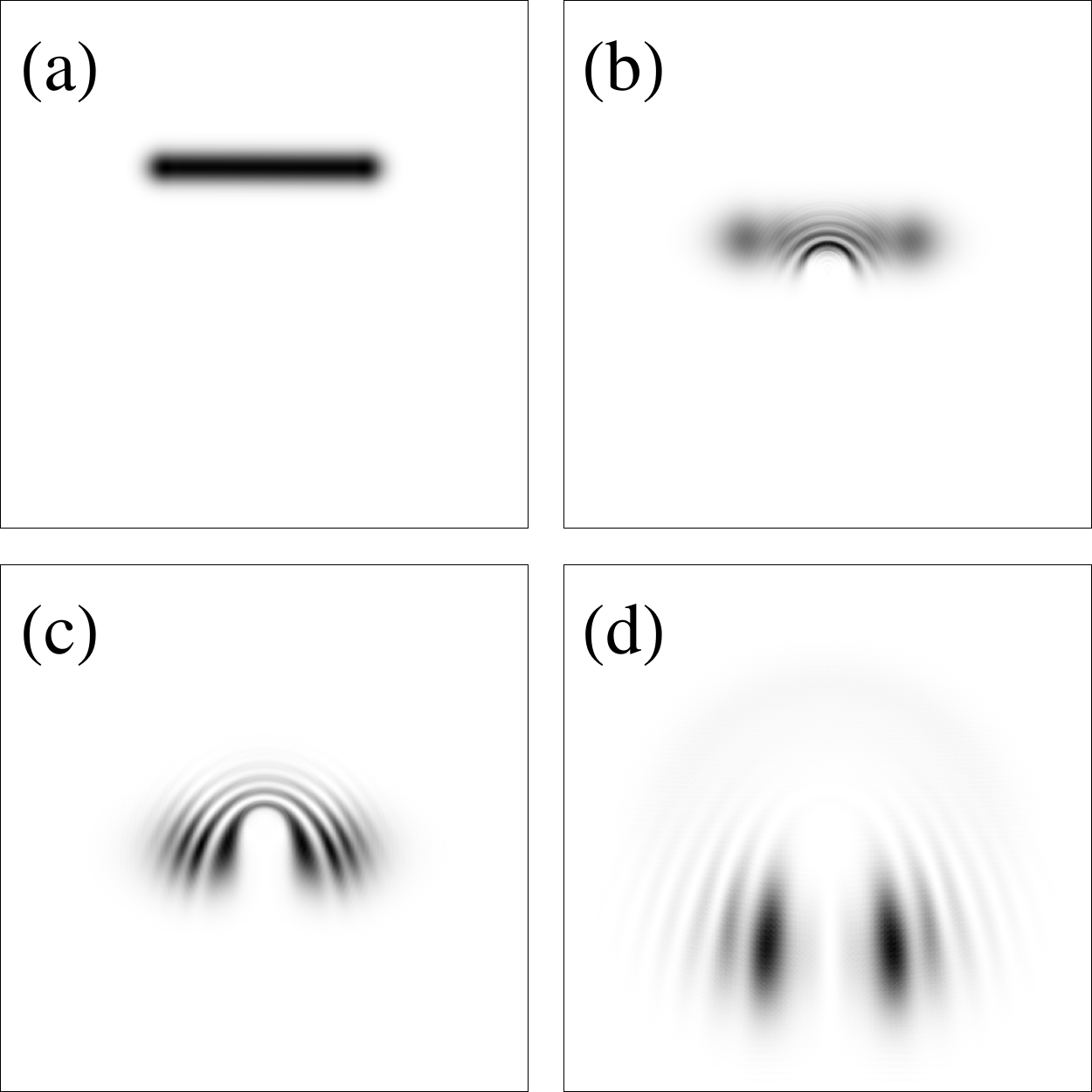}
\caption{\label{fig:fig4}  Probability density plots of a vibronic packet
 being scattering by a conical intersection (not shown) in the ground BO state. Panel (a) 
 shows initial wavepacket at $t=t_{0}$. Panels (b),(c) show packet incident on a CI that is located at the midpoint of each frame. Panel (d) illustrates scattered packet at $t=t_{f}$. }
\end{figure}
\subsection{Time dependent, multi-channel, propagation of wave packets in the diabatic picture}
In the simulation outlined above, we make use of the split operator method\cite{fleck} in order to propagate a packet that is, initially,
asymptotically removed from the scattering center. We first define a set of dimensionless coordinates $ \xi = x/L, \eta=y/L $, and time parameter
$\tau = \hbar \, t/2 m L^2 $ where $L$ is an arbitrary length scale. The resulting time-dependent coupled equations for the diabatic amplitudes
are
\beq 
  i \frac{  \partial{G_{1}}  } {\partial \tau} = - \Bigl ( \frac{\partial^{2}G_{1} }{ {\partial \xi}^2 } + \frac{\partial^{2}G_{1} }{ {\partial \eta}^2 } \Bigr )
+ {\tilde V}_{11} G_{1} + {\tilde V}_{12} G_{2} =0 \nonumber \\
  i \frac{  \partial{G_{2}}  } {\partial \tau} = - \Bigl ( \frac{\partial^{2}G_{2} }{ {\partial \xi}^2 } + \frac{\partial^{2}G_{2} }{ {\partial \eta}^2 } \Bigr )
+ {\tilde V}_{21} G_{1} + {\tilde V}_{22} G_{2} =0 
\nonumber \\
\label{2a.1}
\eeq
where the rescaled couplings ${\tilde V}_{ij}$ are given by
\beq
&& {\tilde V}_{11} = - {\tilde V}_{22}  =\Xi({\tilde \rho}, \tilde \rho_{0})  \, {\tilde \Delta} \, \xi   \nonumber \\
&& {\tilde V}_{12} =  {\tilde V}_{21}  =\Xi({\tilde \rho}, \tilde \rho_{0})  \, {\tilde \Delta} \, \eta 
\label{2a.2}
\eeq 
and
\beq 
&& {\tilde \Delta} = \frac{2 m L^{2} \Delta}{\hbar^{2}} \nonumber \\
&& {\tilde \rho}= \sqrt{\xi^{2} + \eta^{2}}
\label{2a.3}
\eeq
are dimensionless parameters.
We express the diabatic amplitudes in matrix form
\beq
G(\tau) \equiv  \left ( \begin{array}{c}  G_{1}(\tau) \\ G_{2}(\tau) \end{array} \right )
\label{2a.4}
\eeq
and apply the split-operator propagation algorithm\cite{fleck}
\beq
G(\tau+\delta \tau) = U_{KE} \, U_{V} \, U_{KE} \, G(\tau) 
\label{2a.5}
\eeq
where
\beq
U_{KE} = \exp(i \, \frac{\delta \tau}{2} \, \Bigl (\frac{\partial^{2} }{ {\partial \xi}^2 } + \frac{\partial^{2} }{ {\partial \eta}^2 } \Bigr ))
 \left ( \begin{array}{cc}  1 & 0 \\
0 & 1 
\end{array} \right ) 
\label{2a.6}
\eeq
and
\beq
&& U_{V}= \exp(- i \, \delta \tau \, H'_{ad}(x,y)) =  \left ( \begin{array}{cc} U_{11} & U_{12} \\ U_{21} & U_{22} \end{array} \right )
\nonumber \\
\nonumber \\
&& U_{11} =  \cos(\Delta(\xi,\eta) \delta \tau) - i \cos(\phi) \sin( \Delta(\xi,\eta) \delta \tau)  \nonumber \\
&& U_{12} = U_{21} =  -i \sin\phi \, \sin(\Delta(\xi,\eta) \delta \tau )                            \nonumber \\
&& U_{22} = \cos(\Delta(\xi,\eta) \delta \tau) + i \cos(\phi) \sin( \Delta(\xi,\eta) \delta \tau ).  \nonumber \\
\label{2a.7}
\eeq
Here $ 2 \Delta(\xi,\eta) $ is the energy defect between the two BO surfaces shown in Fig. (\ref{fig:fig4}), 
\beq
\Delta(\xi,\eta) \equiv \Delta \quad  {\rm for} \,  {\tilde{\rho} > \tilde{\rho_{0}} }; 
\quad \frac{\Delta}{\tilde \rho_{0}} \, \sqrt {\xi^2+\eta^{2}}  \,  {\rm for} \,  {\tilde{\rho} <= \tilde{\rho_{0}} } \nonumber \\
\label{2a.8}
\eeq
where $\phi$ is the pseudo-angle and $\tan \phi=\eta/\xi$.

With the repeated application of the propagation algorithm Eq. (\ref{2a.5}),  the amplitude $G(\tau)$ at 
$ \tau > \tau_{0}$ is obtained.  The initial packet $\psi_{0}(\tau_{0}) $ is chosen to be the finite slab shown in panel (a) of Figure (\ref{fig:fig4}).
It is incident on the ground adiabatic branch and so we perform a transformation into
the diabatic picture,
\beq 
&& G(\tau_{0}) = U_{c}^{\dag}(\phi) F(\tau_{0})  \nonumber \\
&& \nonumber \\
&& F(\tau_{0}) = \left( \begin{array}{c}  0 \\ \psi_{0}(\tau_{0})  \end{array} \right ).
\label{2a.9}
\eeq
The parameters characterizing $\psi_{0}$ are chosen so that the probability density is appreciable only in the asymptotic  region  where the
ground BO surface has the flat landscape shown in that figure.  It's initial velocity along the positive $\xi$ direction allows it to  proceed toward the origin. 
It encounters the conical intersection near the origin, shown by panels (b),(c) at interim values  $ \tau_{0} < \tau  < \tau_{f}$.  At $\tau_{f}$ the, scattered, packet continues into the, post collision, asymptotic ground BO landscape. 
\begin{figure}[H]
\centering
\includegraphics[width=0.6 \linewidth]{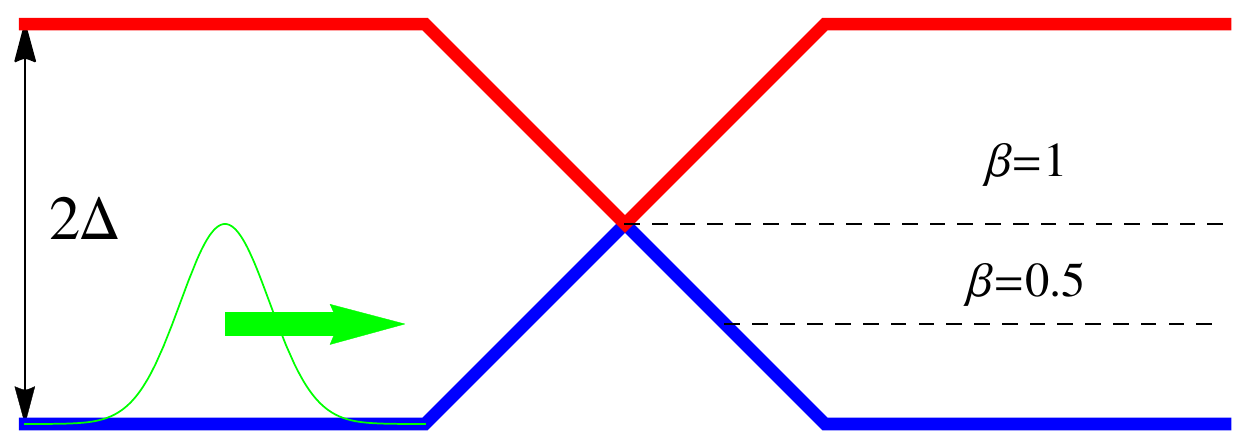}
\caption{\label{fig:fig5} (Color online). Schematic illustration of a packet impinging on a conical intersection for different values of $\beta$.
The diagram (not drawn to scale) is a cross section intersecting the CI with an azimuthal symmetry plane. The blue line is the ground BO
branch, whereas the red line represents the excited state branch.}
\end{figure}
Figure (\ref{fig:fig5}) illustrates the role of the ratio $\beta \equiv k^{2}/\Delta$, where $k^{2}$  is the incident collision energy 
and  $\Delta$ the BO energy defect parameter for the asymptotic region. $\beta=1 $ is the value in which the collision energy 
is equal to the height of the ground branch of the conical intersection at the origin. For $\beta=1/2$ the packet energy is not sufficient
to tunnel through the cone and so diffracts about it. The diffraction peaks are prominent in Figure (\ref{fig:fig4}).
In this study we do not consider the case $\beta >>1$,  for which non-adiabatic
transitions into the excited state are allowed. Our calculations show that for $\beta=1$ a minute fraction $ < 1 \% $ of the initial packet 
suffers a transition. 
\begin{figure}[H]
\centering
\includegraphics[width=0.7 \linewidth]{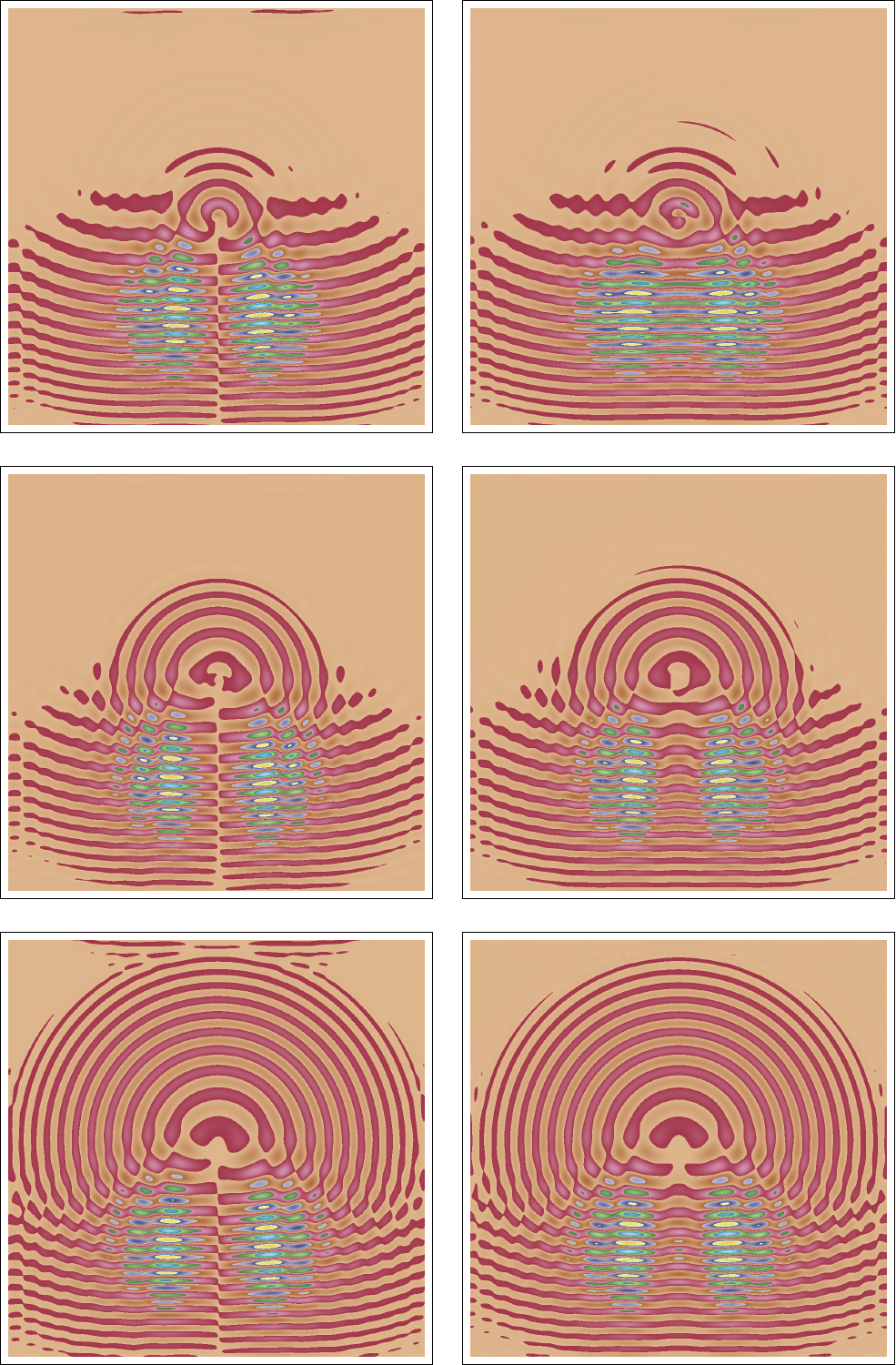}
\caption{\label{fig:fig6} (Color online). Density plots of the imaginary part of the packet amplitude at time $ t=t_{f}$. Initial packet (not shown) at $t_{0}$ is incident at the top of a frame and propagates toward the CI located at the mid-point of each panel.  The row entries
correspond to different values of ratio $\beta = k^{2}/\Delta$.  Top row corresponds to $\beta =1$, middle row to $\beta=1/2$, and
third row to $\beta=1/8$. The left column illustrates solutions to Eq. (\ref{2.7}) in which the adiabatic Hamiltonian is
given by Eq. (\ref{2.8}),  whereas
the right column plots the corresponding solutions for the latter defined in Eq. (\ref{2aa.11}). }
\end{figure}

In Figure (\ref{fig:fig6}) we plot the imaginary part of the amplitude at the post collision time $\tau_{f}$ in the $\xi,\eta$ plane for the
values of $\beta=1, 1/2, 1/8 $.  The top row, and first column, of that figure illustrates the calculated amplitude for the value $\beta=1$.
The subsequent rows, in the first column, correspond to the cases $\beta=1/2$ and $\beta=1/8$ respectively. 
Those panels illustrate that as $\beta$ decreases, back-scattering is enhanced. This behavior
follows from the fact that smaller $\beta$ imply classical turning points near the base and broader sections of the cone (see Fig. (\ref{fig:fig5})). 

Having developed the theory in the diabatic gauge, we consider the coupled equations for the amplitudes $F({\bm R})$ in the adiabatic picture, or gauge. 
We express the adiabatic Hamiltonian,  defined  in Eq. (\ref{2.8}),  as
\beq
H'_{ad}(x,y) = U_{c}(\phi) \, H'_{BO} \, U^{\dag}_{c}(\phi) 
\label{2a.10.1}
\eeq
where $H'_{BO}$ is the diagonal matrix whose eigenvalues are given by Eq. (\ref{2.9})
and $U_{c}$ is defined in Eq. (\ref{2.4}). Proceeding with the PSS expansion,
we arrive at Eq. (\ref{2.5}), with the non-Abelian gauge potential ${\bm A}$ given
by Eq. (\ref{2.6}),  and the BO diagonal matrix now given by $H'_{BO}$. 
We perform a BO projection of it,  onto the ground adiabatic state to get
\beq 
-\, \frac{\hbar^{2}}{2 \mu} \Bigl ( {\bm \nabla} - i {\bm A}_{P} \Bigr )^{2} F_{g}({\bm R}) + 
 {\tilde V}_{g}({\bm R}) F_{g}({\bm R}) = E F_{g}({\bm R}) \nonumber \\
\label{2a.10}
\eeq
where
\beq
{\bm A}_{P} = Tr (P {\bm A} P) = \frac{\bm {\hat \phi}}{2 \rho }
\label{2a.11}
\eeq
is the projected component of the non-Abelian gauge potential ${\bm A}$ and $P$ is a projection operator.
$F_{g}(x,y)$ is the ground adiabatic amplitude and $ {\tilde V}_{g}({\bm R})$  is the sum of the ground BO energy,  given in Eq. (\ref{2.9}),
and the non-adiabatic scalar correction\cite{zyg86,zyg87a,zyg90} $\hbar^{2}/(2 \mu \rho^{2}) $. 
Because we chose collision energies smaller than the energy defect $2 \Delta$, we expect that the projected
BO equation (\ref{2a.10}) provides a good approximation to the amplitude obtained from solution of the fully coupled problem. 
In order to test this hypothesis we introduce a new adiabatic Hamiltonian
\beq
H''_{ad}(x,y) = \Xi (\rho,\rho_{0})  \, \left ( \begin{array}{cc}
   x &  y \exp(-i \phi)  \\
    y \exp(i \phi)  &  -x  
\end{array} \right )
\label{2aa.11}
\eeq
which is a generalization\cite{zwanzig87} of $H'_{ad}$ defined in Eq. (\ref{2.8}). 
 $H''_{ad}$ shares identical potential surfaces to those predicted by $H'_{ad}$ and given
in Eq. (\ref{2.9}). 
Indeed we find
\beq
 H''_{ad}(x,y) = U_{d} H'_{BO} U^{\dag}_{d} \nonumber 
 \eeq
 \beq
 U_{d} = 
 \left(
\begin{array}{cc}
 e^{\frac{1}{2} i \sin (\phi )-\frac{i \phi }{2}} \cos \left(\frac{\phi
   }{2}\right) & -e^{-\frac{i \phi }{2}-\frac{1}{2} i \sin (\phi )} \sin
   \left(\frac{\phi }{2}\right) \\
 e^{\frac{i \phi }{2}+\frac{1}{2} i \sin (\phi )} \sin \left(\frac{\phi
   }{2}\right) &
 e^{-\frac{1}{2} i \sin (\phi )+\frac{i \phi }{2}} \cos \left(\frac{\phi
   }{2}\right)
 \\
\end{array}
\right). \nonumber 
% \label{2a.12}
\eeq
$H''_{ad}$  differs from $H'_{ad}$ as it embeds $H'_{BO}$ with unitary operator $U_{d}$ instead of $U_{c}$. It is also
evident that $U_{d}$, like $U_{c}$, is single-valued.
With it we 
 obtain the PSS equation (\ref{2.5}) with $V({\bm R}) = H'_{BO}(x,y)$  and 
\beq
 {\bm A} = iU_{d}^{\dag} {\bm \nabla} U_{d}= \nonumber 
 \eeq
 \beq
 -\, \frac{{\bm {\hat} \phi}}{2 \rho}  \left(
\begin{array}{cc}
 0 & e^{-i \sin (\phi )} (\sin (\phi )+i) \\
 e^{i \sin (\phi )} (\sin (\phi )-i) & 0 \\
\end{array}
\right).  \nonumber
% \label{2a.13}
\eeq
Because the diagonal components of this matrix vanish, the BO projection of the PSS equations
again leads to Eq. (\ref{2a.10}) but with the important difference that ${\bm A}_{P}=0$. In addition, the scalar non-adiabatic
correction is modified to $\hbar^{2}(1+\sin^{2}(\phi))/(8 \mu \rho^{2}) $. According to the BO approximation,
 and the fact that ${\bm A}_{P}=0$, $H''_{ad}$ does not allow a molecular AB effect 
 despite the fact that $H''_{ad}$ shares an identical CI to that predicted by $H'_{ad}$. 

We propagate the packet in the
diabatic picture,  as described above, but replace $H'_{ad}$ with $H''_{ad}$. The results of those simulations are
shown in the second column of Fig. (\ref{fig:fig6}).
In that figure, the rows correspond to results of
calculations using the respective values for $\beta$ itemized above.  Note that the overall structure of the 
calculated amplitudes is similar to that shown by
the panels in the first column.
They both show similar  trends in the backscattering amplitude as $\beta$ varies,  and validates the 
BO approximation in that the gross features of the elastic scattering amplitude are correctly predicted by
the ground state BO potential $V_{BO}=-\Delta/\rho_{0} \, \sqrt{x^2+y^2} $ in the region $ \rho<\rho_{0}$.
However,  the amplitudes illustrated in the left column do not exactly match 
those on the right. The former amplitudes exhibit a prominent phase dislocation line that starts at the origin
and proceeds along the positive $x$ axis. This feature is absent in the corresponding panels of the second column
in Figure (\ref{fig:fig6}) and appears to be independent of the collision parameter $\beta$. 
In appendix A it is shown how topological phase dislocation lines are a consequence of the induced
Abelian vector potential ${\bm A}_{P} $ that is explicit in the adiabatic gauge (representation).
\begin{figure}[ht]
\centering
\includegraphics[width=0.9 \linewidth]{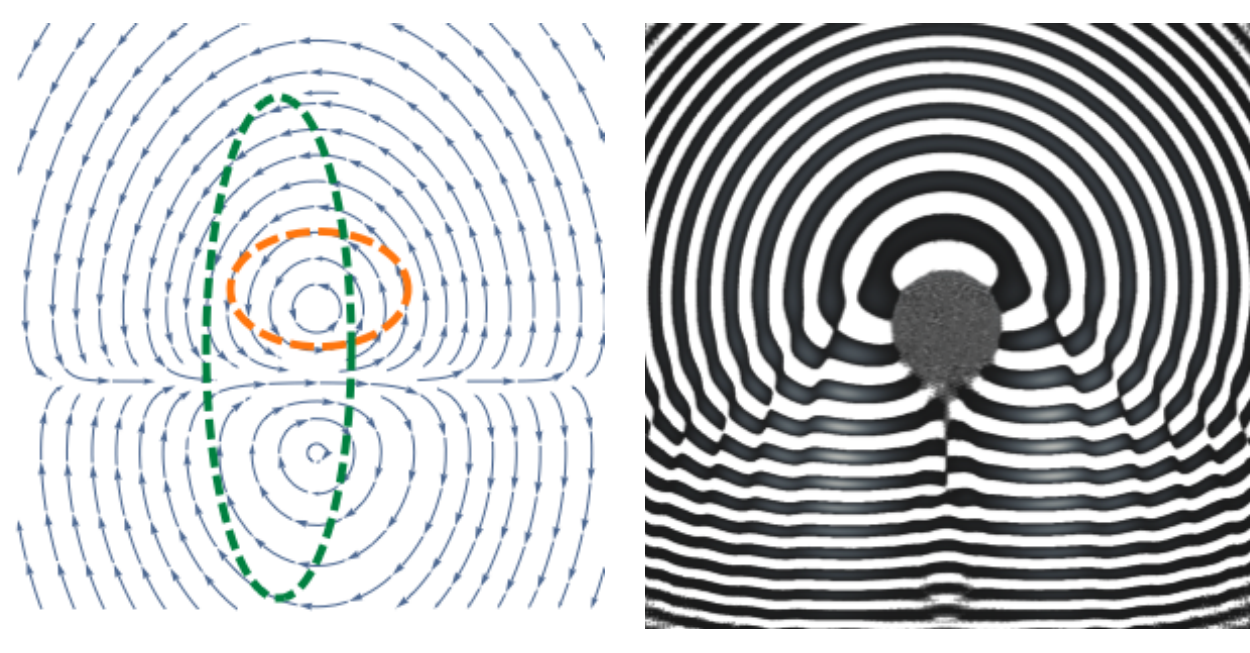}
\caption{\label{fig:fig7} (Color online). Left panel: Flow maps of the vector potential, that describe a pair of CI's, given in
Eq. (\ref{3.03}). Dashed lines represent closed paths that define Wilson loop integrals. Right panel: Complex component of
vibronic amplitude, as it is scattered by the pair of CI's located at the origin, and a point downwind from the scattering center.
The dark circular shaded region centered at the origin represents an impenetrable obstacle. The phase dislocation line
terminating at the point $ \xi=0, \eta_{0}=3$ is clearly visible in the figure.}
\end{figure}
\section{Summary and conclusion}
We introduced a two-state molecular collision model which features a conical  intersection that asymptotically
correlates to a pair of BO states separated by a finite energy gap $2 \Delta$. For it, we numerically solved the time dependent
Schroedinger equation in the diabatic gauge (representation).  An
incident wavepacket of mean energy, less than the energy gap, was chosen and propagated in time as it was scattered by the conical
intersection. The global phase structure
of the vibronic amplitudes was analyzed and found to exhibit an underlying topological order associated with
phase dislocation lines identical to ones observed in standard AB scattering. The dislocation lines, suggesting fractional
topological charge, are not affected by variation in the total collision energy as long the inequality  $k^{2}/2\mu < 2 \Delta$ is satisfied.
We conclude that this feature demonstrates persistence of topological behavior beyond the strictly adiabatic limit. 
It differs from the conclusions of a previous study\cite{gross15}  that argues survival of the MAB effect only in the limit where  $\mu \rightarrow \infty$.
The results of our numerical study, that does not rely on the BO approximation,  confirms the fidelity  of the latter in applications for 
realistic molecular collision systems provided
that a gauge vector potential is minimally coupled to the vibronic amplitude in the adiabatic gauge.

There have been numerous  efforts over the decades to predict observable MAB behavior in reactive scattering settings,
but, as of yet, there exist no compelling evidence of the latter in reports of laboratory measurements. One explanation that is offered
for the lack of
a MAB  ``smoking gun'' is the possibility of cancellation of partial-wave phase shifts generated by the CI scalar potential with that produced by the Mead-Truhlar vector potential.  
A recent study\cite{bala15} argued that this cancellation effect might be circumvented
in ultra-cold molecular collisions where s-wave scattering dominates. According to Wigner threshold theory, phase
shifts generated by a short range potential scale as a power of the incident collision wave number and therefore the partial wave phase shifts
$\delta(k) \rightarrow 0$ as $ k \rightarrow 0$. In the ultra-cold limit the leading order contribution to the cross section is
generated by the isotropic s-wave.
However, the presence of a long range vector potential $ {\bm A}_{MAB}$ predicts phase shifts $|\delta_{MAB}|_{m} = \pi/2 $ 
for the m'th partial wave \cite{Au84,hen80}.  In pure AB scattering,
the leading order term that survives in the $k R \rightarrow 0$ limit is (see Appendix A) proportional to
\beq  J_{1/2}(k R) (1 + \exp(i \phi) )
\label{3.01}
 \eeq
and differs from standard 2D s-wave behavior by the fact that the radial term $J_{1/2}$ vanishes at the origin, and both the
$s$ and $p$-waves contribute identical phase shifts. Also, in contrast to short range potentials in which only s-waves
survive in the ultra-cold cold limit,  Eq. (\ref{3.01} ) 
predicts anisotropic scattering. 
Here we propose an additional mechanism by which AB phase shifts at larger impact parameters may be suppressed.
If ${\bm A}_{MAB}$ is screened at larger
internuclear distances due to multiple CI's, does the MAB effect arise in the ultra-cold limit ?  
In order to address this question, within the scope of our 2-state model, we need to calculate the gauge
structure of a general electronic (2-state) model Hamiltonian
\beq
H_{ad}(x,y) = \left ( \begin{array}{cc} h(x,y) & g(x,y) \\ 
                  g(x,y) & -h(x,y) \end{array} \right ).
\label{3.02}
\eeq
The regular, real-valued, functions $h(x,y),g(x,y)$ were chosen (see appendix B) so that $H_{ad}$ 
possesses  a pair of conical intersections located at the origin and at a point downwind $(x=x_{0}, y=0) $ from the scattering center.
In addition to the conical intersection potential surface, 
we added a radial symmetric barrier centered at the origin in order to define a scattering center.
In right the hand panel of Fig. (\ref{fig:fig7}) we plot the imaginary part of the vibronic amplitude in order to uncover its global
phase structure. Clearly evident in this figure is a phase dislocation line, starting at the origin (not show because tunneling
into the radial barrier is prevented) and extending along the vertical
scattering axis. However, unlike the case
discussed in the previous sections, the phase dislocation line terminates at the location of the second conical intersection.
According to the results of Appendix B the gauge potentials associated with this pair of CI's is
\beq
&& A_{x} = \frac{y(2 x-x_{0})}{2(y^{2} + x^{2}(x_{0}-x)^{2} )}  \nonumber \\
&& A_{y} = \frac{x(x_{0}-x)}{2(y^{2} + x^{2}(x_{0}-x)^{2})  }. 
\label{3.03}
\eeq
In the left-most panel of Fig. (\ref{fig:fig7}) we plot streamlines for vector potential ${\bm A}$ whose components
are given by Eq. (\ref{3.03}). Superimposed, in this figure,
 are closed paths $ C_{i}$ for which we evaluate the Abelian Wilson loop integral for ${\bm A}$ i.e.
$$ \exp(i \oint_{C_{i}} {\bm A} \cdot d{\bm R} ). $$
For the loop that encloses only a single CI, shown by the orange line in that figure,
the integral has the value $-1$, in contrast to value obtained, $+1$, for the (green) loop that encloses both CI's.
 In the latter, topological holonomy is absent. Our observation is in harmony with the conclusion of
 previous investigations of systems
possessing a CI pair \cite{yark99}. How this effective screening at larger impact parameters affects scattering
properties, at both the ultra-cold and high energy limits is under current investigation. 
In addition, we pose the question: what
happens when the collision energy is much larger than the energy defect $\Delta$ between the
ground and excited electronic BO surfaces. Evidence gleaned from recent investigations concerning geometric phase effects 
in two-state systems\cite{zyg12,zyg15},
as well as in spinor Bose-Fermi mixtures\cite{phuc15}, suggests a transition from Abelian to non-Abelian behavior and in the limit where $ k^{2}/2\mu >> \Delta$ decoupling to a null
effect.  If and how this Abelian to non-Abelian crossover occurs in molecular systems is an open question and
deserves closer examination. 

\begin{acknowledgments}
I wish to acknowledge support by the NSCEE for use of the Intel Cherry-Creek computing cluster.
\end{acknowledgments}

\appendix
\section{Topological charge of $\Psi_{AB}$  } 
Consider a complex scalar field in two dimensions and express it in the form $\psi(x,y) = \rho(x,y) \exp(i \chi(x,y)) $ where $\rho(x,y) \geq 0$ and
$ \chi $ are real functions. The quantity 
\beq
S_{C} \equiv \frac{1}{2 \pi}  \oint_{C} d {\bm R} \cdot {\bm \nabla } \chi,  
\label{a0.1}
\eeq
where $C$ is a closed loop (not crossing the zeros of $ \rho$ ) is called the topological charge\cite{dennis01,Berry80}.
 Below we offer an illustration
of  this property in solutions of the Schroedinger equation that is  minimally coupled
to the vector potential ${\bm A}_{MAB}$.  

Consider the partial wave solutions of the AB equation\cite{ab59}
\beq
 J_{|m-\alpha|}(k r ) \, \exp(i m \phi)
\label{a0.2}
\eeq
where $ m$ is an integer.
They possess nodal lines at the zeros of $ J_{ |m-\alpha |} $ and in the annular region between two nodes,  $ \chi= m \, \phi + \phi_{0} $ where
the constant $ \phi_{0} = 0,\pi $ corresponds to the cases where $ J_{| m-\alpha |}$ is positive and negative respectively. Thus
 $ S_{C} $ in a given region (not crossing the nodal lines) has integer topological charge $m$.  For
$ \alpha=1/2$ we construct the  linear combination  of the $m=0,1$ partial waves
\beq
&& {\tilde \psi}(r,\phi)  \equiv J_{|0-1/2|}(k r) + J_{|1-1/2|}( k r) \exp(i \phi) = \nonumber \\
&& J_{1/2}(k r) \Bigl ( 1+\exp(i \phi) \Bigr ),
\label{a0.3}
\eeq
which is proportional to the leading order term in the $ k r \rightarrow 0 $ limit of $\psi_{AB}$.
In addition to the concentric nodal line structure described above, $\tilde \psi$  vanishes along the negative $x$ axis ($\phi = \pi$) 
and we find $ \chi = \phi/2 \pm \phi_{0}$ in a given annular region. Because  $\chi $ is not defined on a nodal line, we allow
loops $C$ in a given annular region that start at $\phi=-\pi$ and end at $\phi=\pi$.
 Now, using Eq. (\ref{a0.3}),
\beq
{\bm \nabla }\chi = Im \, ( \frac{ {\tilde \psi}^{\dag} {\bm \nabla}{\tilde \psi} }{ {\tilde \psi}^{\dag} {\tilde \psi} } )
=  \frac{\bm {\hat \phi}}{r} \,  Im \Bigl (\frac{\exp(i\phi)}{ \, (\exp(i\phi)+1) }  \Bigr )
\nonumber
\eeq
and so the topological charge
\beq
S_{C} = \frac{1}{2\pi } \, \int_{-\pi}^{\pi} d\phi \,\frac{1}{2} 
\nonumber
\eeq
has the fractional value $1/2$. 
It is topological in the sense that the phase discontinuity along the negative $x$-axis persists
regardless of the wavenumber $k$, and, as we demonstrate below, the nature of the short range potential.

In the scattering region, the most general solution can be expressed as a partial wave
expansion
\beq
&& \psi(r,\phi) = \sum_{m=-\infty}^{\infty}  \, \, exp(i m \phi) \times  \nonumber \\
&& ( a_{m} J_{| \alpha-m |}(k r) + 
b_{m}  H^{(1)}_{| \alpha-m |}(k r) )
\label{a0.4}
\eeq
where the coefficients $a_{m},b_{m}$ are uniquely determined by the asymptotic boundary conditions and the requirement
that Eq. (\ref{a0.4}) match the logarithmic derivatives of the $m'th$ radial partial wave at some matching distance $r_{c}$.
The radial wavefunctions $R_{m}(r)$ obey
\beq
&& R''_{m}(r) +\frac{R'_{m}(r)}{r} - \frac{(m-\alpha)^2}{r^{2}} R_{m}(r) + \nonumber \\
&& (k^2-U(r)) R_{m}(r) =0 
\label{a0.5}
\eeq
and for $\alpha=1/2$ we find that $R_{m}$ is identical for values where $ |m-1/2| = (2 n+1)/2$ and $n=0,1,2...$.
We construct the logarithmic derivative 
\beq
y_{m} \equiv k \,R_{m}'(k r_{c})/R_{m}(k r_{c}) 
\eeq
at the matching distance. $y_{m}$, is also invariant for $m$ that satify $ |m-1/2| = (2 n+1)/2$.
As we require the incoming wave to have the form $\exp(-i k r \cos(\phi)) \exp(i \phi/2) $ \cite{ab59}
we find that
\beq
&& a_{m} = (-i)^{|m-\alpha|}  \nonumber \\
&& b_{m} = (-i)^{|m-\alpha|} \frac{ y_{m} J_{|m-\alpha|} - J'_{|m-\alpha|} }{ H'^{(1)}_{|m-\alpha|}  - y_{m} H^{(1)}_{|m-\alpha|} }.
\label{a0.5a}
\eeq
On the r.h.s of this equation the Bessel functions, and their derivatives, are evaluated at $k r_{c}$.
For $ \alpha=1/2 $ we note that
\beq
&& a_{-n} = a_{n+1}  \nonumber \\
&& b_{-n} =b_{n+1} 
\label{a0.6}
\eeq
where $ n=0,1,2,...$. Using this relation we can 
re-write Eq. (\ref{a0.4}) in the form
\beq
&& \sum_{n=1}^{\infty} \exp(i n \phi) \Phi_{n}(\phi) \times \nonumber \\
&& \Bigl [ a_{n} J_{n-1/2}(k r) + b_{n} \, H^{(1)}_{n-1/2} (k r) \Bigr  ] \nonumber \\
&&  \Phi_{n}(\phi) \equiv  1+ \exp(-i (2 n+1) \phi) .
\label{a0.7}
\eeq
Because $ \Phi_{n}(\pm \pi) $ vanishes identically and is an odd function of $\phi$ about $\phi=\pm\pi$ for all $n$, $\psi(r,\phi)$
suffers a phase discontinuity along this line. The phase dislocation shows prominently in the illustrations in Fig (\ref{fig:fig2}).
It is independent of both the collision energy (or wavenumber $k$) and the parameters that define the short range potential.

\section{Multiple conical intersections}

Consider the Hamiltonian
\beq
H(x,y) = \left ( \begin{array}{cc} h(x,y) & g(x,y) \\ 
                  g(x,y) & -h(x,y) \end{array} \right )
\label{b1}
\eeq
where $h(x,y) \equiv h, g(x,y) \equiv g$ are continuous real-valued functions defined on the 2D $x,y$ plane. We consider only those functions in which
$H(x,y)$ possess a finite set of discrete zeros. 
The eigenvalues of $H$ are $ E = \pm \sqrt{g^2+h^2} $. The square root symbol refers
to the positive branch. Expressing the eigenvectors in matrix form
\beq
\psi \equiv  \left( \begin{array}{c} c_{1} \\ c_{2} \end{array} \right )
\label{b2}
\eeq
we require, for the lower energy eigenvalue $E_{-}=- \sqrt{g^2+h^2}$ 
\beq
c_{1} ( h + \sqrt{g^2+h^2} ) + c_{2} \, g =0 
\label{b3}
\eeq
and find a possible (unnormalized) real eigenvector for $ E_{-}$ 
\beq
{\tilde \psi}_{a} = \left( \begin{array}{c} h-\sqrt{g^2+h^2} \\ g  \end{array} \right )                 		  
\label{b4}
\eeq
Because $h,g$ are regular on the $x,y$ plane so is ${\tilde \psi}_{a}$, however
if there are lines ${ \cal C}$ in the $x,y$ plane in which $g(x,y)=0$ then $\psi_{a}$ cannot be normalized
in the regions where $ h(x,y) \ge |h(x,y)| \cap {\cal C} $. 

We therefore define a new (real-valued) eigenvector
\beq
{\tilde \psi}_{b} = \left( \begin{array}{c} g \\  -h-\sqrt{g^2+h^2}  \end{array} \right )                 		  
\label{b5}
\eeq
which does not vanish in the region where ${\tilde \psi}_{a}=0$. ${\tilde \psi}_{a}, {\tilde \psi}_{b} $
are not linearly independent
in the regions where they both have positive norm. However we can define
the complex-valued eigenstate
\beq
&& \psi_{-} \equiv \psi_{a}(x,y) + i \, \psi_{b}(x,y) \nonumber \\
&&   \psi_{a} \equiv \frac{{\tilde \psi}_{a}}{2 \sqrt{g^2+h^2}} \nonumber \\
&&   \psi_{b} \equiv \frac{{\tilde \psi}_{b}}{2 \sqrt{g^2+h^2}}. \nonumber
\label{b6}
\eeq
$ \psi_{-} $ is normalized to unity and, with the exception of isolated points $P_{i} =(x_{i},y_{i})$ in which
 $h(P_{i})$ and $g(P_{i})$ vanish identically, it is well behaved everywhere.  

In the same manner we obtain the normalized complex-valued eigenvector $\psi_{+}$ for the branch $E_{+} = \sqrt{g^2+h^2}$.
\beq
&& \psi_{+} \equiv \psi_{c}(x,y) + i \, \psi_{d}(x,y) \nonumber \\
&&   \psi_{c} \equiv \frac{{\tilde \psi}_{c}}{2 \sqrt{g^2+h^2}} \quad
\psi_{d} \equiv \frac{{\tilde \psi}_{d}}{2 \sqrt{g^2+h^2}}. \nonumber \\
&& {\tilde \psi}_{c} = \left( \begin{array}{c} h+\sqrt{g^2+h^2} \\ g \end{array} \right ) \nonumber \\ 
&& {\tilde \psi}_{d} = \left( \begin{array}{c} g \\  -h+\sqrt{g^2+h^2}  \end{array} \right )
\label{b7}
\eeq
From these solutions we can construct the unitary matrix 
\begin{widetext}
\beq
U \equiv \frac{1}{2 \sqrt{g^2+h^2} } \, \left( \begin{array}{cc}  h+\sqrt{g^2+h^2} + i\, g & -g \,  -i  (  h - \sqrt{g^2+h^2})  \\
                                  g +  i \, (  -h+\sqrt{g^2+h^2}) & -i g +  h+\sqrt{g^2+h^2}   \end{array} \right )
\label{b8}
\eeq
\end{widetext}
It is single-valued everywhere except at the points $P_{i}$ defined above. $U$ diagonalizes Hamiltonian Eq.(\ref{b1}) so
that
\beq
&& H= U^{\dag} H_{BO} U  \nonumber \\
{\rm where}  \nonumber \\
&&  H_{BO} \equiv \left( \begin{array}{cc} \sqrt{h^2+g^2} & 0 \\ 0 & - \, \sqrt{h^2+g^2} \end{array} \right ). 
\label{b9}
\eeq
For the special case $ h(x,y)=x , g(x,y) = y$ we obtain the unitary operator defined in Eq. (\ref{2.4}). From $U$ we obtain
the non-Abelian gauge potential 
\beq
{\bm A} \equiv i \, U^{\dag} {\bm \nabla} U. 
\label{b10}
\eeq
It is a pure gauge\cite{zyg15,comment1} in the sense that the Wilson loop integral satisfies the identity
\beq
P \exp(i \int_{C} {\bm A} \cdot d {\bm R} ) = {\bm I}, 
\label{b11}
\eeq
where $C$ is any closed loop that does not intersect $P_{i}$ in the $x,y$ plane, ${\bm I}$ the unit matrix, and  $P$
refers to a path-ordered integral\cite{zyg15}. However the projected gauge potential 
\beq
{\bm A}_{g} \equiv Tr \, P_{g} {\bm A} P_{g} 
\label{b12}
\eeq
may be non-trivial and  necessarily not satisfy Eq. (\ref{b11}). Here $P_{g}$ is a projection operator into the ground state of $H_{BO}$.

Using definitions Eqs. (\ref{b8}), (\ref{b10}) and Eq. (\ref{b12}) we find
\beq
{\bm A}_{g} = \frac{h {\bm \nabla} g - g {\bm \nabla} h}{2 (g^2+h^2) }.
\label{b13}
\eeq
Evaluating the curvature $ {\bm H} = {\bm \nabla} \times {\bm A}_{g} $ we find that it vanishes identically except possibly
at the points $P_{i}$ where ${\bm A}_{g}$ is singular. Suppose
a set $n$ of such points $P_{1},P_{2} .. P_{n}$ exist. We now construct a line integral
\beq
\int_{C} d{\bm r} \cdot {\bm A}_{g}
\label{b14a}
\eeq
where $C$ is a circular contour of radius $R$ centered at the origin. According to Stokes theorem, and the fact that
$ {\bm H}=0$ in the region not including these points, we find
\beq
\int_{C} d{\bm r} \cdot {\bm A}_{g} = \sum_{i}^{n} \int_{C_{i}} d{\bm r} \cdot {\bm A}_{g} 
\label{b14}
\eeq
where $C_{i}$ is the infinitesimal contour surrounding the point $P_{i}$.  If integration along $C$ is counterclockwise
so it is with the paths $C_{i}$.

Let us consider the line integral surrounding the point $P_{0}=(x_{0},y_{0})$. We define the
coordinates $x'= \rho \cos\phi, \, y'= \rho \sin\phi $ where $ \rho $ is the distance
from a point $x,y$ on the path $C_{0}$ to $P_{0}$ and $\phi$ is the angle of a line from $P_{0}$ to
a point on $C_{0}$ makes with the $x'$ axis. We then get
\beq
\int_{C_{0}} d{\bm r} \cdot {\bm A}_{g} = \int_{0}^{2\pi} d\phi \, \Bigl (
\frac{ h \, \partial_{\phi} g - g \, \partial_{\phi} h  }{2 (g^{2} + h^{2})} \Bigr) .
\label{b15}
\eeq
Because we can shrink $C_{0}$ arbitrarily close to $P_{0}$  we can express $h,g$ in that neighborhood as 
\beq
&& h(x',y') \approx  h_{x} \rho \cos\phi + h_{y} \rho \sin\phi  \nonumber \\
&& g(x',y') \approx  g_{x} \rho \cos\phi + g_{y} \rho \sin\phi  
\label{b16}
\eeq
where the constant  
$$ h_{x} \equiv  \frac{\partial h}{\partial x}$$ is evaluated at the point $P_{0}$ and we used the fact that $h(P_{0})=g(P_{0}) = 0$. 
The constants $h_{y},g_{x},g_{y}$ are defined in a similar way. Inserting expressions Eq. (\ref{b16}) into
Eq. (\ref{b15}), we obtain
\begin{widetext}
\beq
&& \int_{C_{0}} d{\bm r} \cdot {\bm A}_{g} = 
 \int_{0}^{2\pi} d\phi \,  \frac{g_{y} h_{x}-h_{y} g_{x}}{2 \Bigl ( (g_{x}^2+ h_{x}^{2})\cos^{2}\phi 
+ (g_{y}^2+ h_{y}^{2})\sin^{2}\phi + (g_{x} g_{y}+ h_{x} h_{y}) \sin2\phi \Bigr ) }.
\eeq
\end{widetext}
The second integral is evaluated to give,
\beq
\int_{C_{0}} d{\bm r} \cdot {\bm A}_{g} = \text{sgn}(g_{y} h_{x}-h_{y} g_{x}) \pi.
\label{b17}
\eeq
Using this result and Eq. (\ref{b14}) we make the following observation for the Abelian Wilson loop integral,
\beq
\exp(i \int_{C} d {\bm r} \cdot {\bm A}_{g})   = \pm 1 
\eeq
for the cases where  $ C$ encloses an even number, or odd, number of singularities respectively (provided that the r.h.s. of Eq. (\ref{b17}) does not vanish).

\bibliographystyle{apsrev4-1}
%\bibliographystyle{naturmag}
%\bibliography{bib2015AB}

\begin{thebibliography}{37}
\bibitem [{com()}]{comment1}%
  \BibitemOpen
  \href@noop {} {}\bibinfo {note} {In some literature, the term pure gauge is
  applied to vector potentials (connections) whose curvature vanishes but which
  may possess, as does the AB vector potential, a non-trivial Wilson loop
  phase. Here, we reserve this terminology for connections that have vanishing
  curvature and whose Wilson phase loop integral is unity. With the latter
  definition, one can always find a non-singular gauge transformation in which
  ${\bm A}=0$.}\BibitemShut {Stop}%
\end{thebibliography}

%merlin.mbs apsrev4-1.bst 2010-07-25 4.21a (PWD, AO, DPC) hacked
%Control: key (0)
%Control: author (72) initials jnrlst
%Control: editor formatted (1) identically to author
%Control: production of article title (-1) disabled
%Control: page (0) single
%Control: year (1) truncated
%Control: production of eprint (0) enabled
%

\end{document}